\documentclass[prd,showpacs,amsmath,showkeys,twocolumn,floatfix,amssymb, preprintnumbers, nofootinbib, superscriptaddress]{revtex4} 
\usepackage{epsfig,dcolumn}
\usepackage{graphicx}
\usepackage{comment} 
\usepackage[usenames]{color}
\usepackage{graphicx}
\usepackage{bm}
 \usepackage{ifpdf}

\def\l{{\bf l}}

\def\lsim{\mathrel{\rlap{\lower4pt\hbox{\hskip1pt$\sim$}}
    \raise1pt\hbox{$<$}}}
\def\gsim{\mathrel{\rlap{\lower4pt\hbox{\hskip1pt$\sim$}}
    \raise1pt\hbox{$>$}}}
\begin{document}

\title{  Coupled-channel scattering on a torus}

\author{Peng~Guo}
\email{pguo@jlab.org}
\affiliation{Thomas Jefferson National Accelerator Facility, 
Newport News, VA 23606, USA}

\author{Jozef~J.~Dudek}
\affiliation{Thomas Jefferson National Accelerator Facility, 
Newport News, VA 23606, USA}
\affiliation{Department of Physics, Old Dominion University, Norfolk, VA 23529, USA}

\author{Robert~G.~Edwards}
\affiliation{Thomas Jefferson National Accelerator Facility, 
Newport News, VA 23606, USA}

\author{Adam~P.~Szczepaniak}
\affiliation{Physics Department, Indiana University, Bloomington, IN 47405, USA}
\affiliation{Center For Exploration  of Energy and Matter, Indiana University, Bloomington, IN 47408, USA.}

\preprint{JLAB-THY-12-1659}

\date{\today}

\begin{abstract} 
 Based on the Hamiltonian formalism approach,    a generalized L\"uscher's formula for two particle scattering in both the elastic and coupled-channel cases in moving frames  is derived from a relativistic Lippmann-Schwinger equation. Some strategies for extracting scattering amplitudes for a coupled-channel system from the discrete finite-volume spectrum are discussed and illustrated with a toy model of two-channel resonant scattering.  This formalism will, in the near future, be used to extract information about hadron scattering from lattice QCD computations.     
  \end{abstract} 

\pacs{12.38.Gc, 11.80.Gw, 11.80.Jy, 13.75.Lb}

\maketitle


\section{Introduction}
\label{intro} 

Hadron spectroscopy in lattice QCD is entering a new era, in particular, recent developments in the application of variational methods \cite{Michael:1985ne,Luscher:1990ck,Blossier:2009kd} to large bases of hadron interpolating fields have made the extraction of the excited spectrum of hadronic states a realistic possibility (see e.g. \cite{Jo:2010,Edwards:2011, Jo_scatt:2012}). Since excited hadrons appear as resonances in the continuous distribution of multi-hadron scattering states, to study hadron spectroscopy one requires evaluation of scattering amplitudes, but because lattice QCD is formulated in Euclidean space, we do not have direct access to these \cite{Maiani:1990ca}. Fortunately, in a finite volume, interactions between particles as they evolve from the {\it in} to the  {\it out} states lead to discrete changes in a free particle's energy that can be related to the scattering amplitude \cite{Lusher:1991}. 

 Various extensions to the framework derived by L\"uscher in \cite{Lusher:1991} have been proposed
 which allow for evaluation outside the center-of-mass frame \cite{Gottlieb:1995,Lin:2001,Christ:2005,Bernard:2007,Bernard:2008},  and to include the coupled-channel effects that can appear above the inelastic threshold \cite{Liu:2005,Doring:2011,Aoki:2011,Briceno:2012yi, Hansen:2012tf}.  The original approach and its extensions to describe the moving center-of-mass frame have been quite successfully used by the lattice community to extract elastic hadron-hadron scattering phase shifts\cite{Aoki:2007rd, Sasaki:2008,Feng:2011,Jo_scatt:2011,Jo_scatt:2012,Beane_scatt:2012,Lang_scatt:2011, Aoki:2011yj}.

In this work, we discuss a generalization of L\"uscher's method for relativistic scattering in terms of a Hamiltonian where the specific interactions considered are based on a relativistic particle exchange model. We apply the technique to a  two-channel system and a generalized L\"uscher's equation for scattering in a moving frame is derived based on the relativistic Lippmann-Schwinger  equation. The coupled-channel system has been considered previously, \cite{Liu:2005,Doring:2011,Briceno:2012yi, Hansen:2012tf}, and our result agrees with these works. A novelty of the present work is to discuss practical strategies for extraction of scattering amplitude parameters from lattice simulations of a coupled-channel system. These strategies are demonstrated using an explicit toy model of resonant two-channel scattering.

The paper is organized as follows. A discussion of elastic scattering in a finite-volume is given   in Section \ref{singlechannel}, with extension to the coupled channel system in Section \ref{coupledchannel}.   Strategies for extracting scattering amplitudes from measured discrete finite-volume spectra are presented in Section \ref{toycoup}. The summary and outlook are given in Section \ref{summary}.


\section{Finite-volume elastic scattering in a Hamiltonian framework} \label{singlechannel}

In this section we present relativistic two-particle scattering on a torus using the Hamiltonian formalism developed in \cite{Glazek:1993,Bakker:2003}. In particular we consider a complex scalar field, $\Phi$, describing a charged boson, $\phi^\pm$, of mass $m$, and its interactions with a neutral boson, $\theta$, which acts as a force carrier and is described by a real scalar field, $\Theta$.  We first  derive the L\"uscher formula describing the finite-volume spectrum of  the  asymptotic two-particle, $\phi^+ \phi^-$ state, 
\begin{equation}
\det \left[ \delta_{JM,J'M'} \cot \delta_J(k) - \mathcal{M}^{(\mathbf{ Q})}_{JM,J'M'}(k) \right] = 0, \nonumber
\end{equation}
where the volume and scattering-frame dependent matrix element $ \mathcal{M}^{(\mathbf{ Q})}_{JM,J'M'}$ is defined in Eq.(\ref{exp_coef}) and  (\ref{mat_element}), and the center-of-mass frame scattering momentum,
 $k$,  is  related to the energy by $\sqrt{s/4 - m^2}$.
The model corresponds to a Lagrangian density, 
\begin{equation} \label{Lagrangian}
\mathcal{L} = \partial_\mu \Phi^* \partial^\mu \Phi - m^2 \Phi^* \Phi  + \tfrac{1}{2}\partial_\mu\Theta \partial^\mu \Theta - \tfrac{1}{2}\mu^2 \Theta^2 - g \Theta \Phi^* \Phi, 
\end{equation}
from which the Hamiltonian can be derived  following the canonical procedure of instant-time quantization (see Appendix \ref{hamiltonian}) \cite{Dirac:1949}. Taking matrix elements of the Hamiltonian in an infinite basis of Fock states spanned by any number of $\phi$ and $\theta$ bosons one can obtain a Schr\"odinger equation 
  $\hat{H} |\Psi\rangle = E  |\Psi\rangle$ for the eigenstates of the theory.  Assuming $\mu \gg m$,  in describing low-energy $\phi$-boson scattering we can  truncate the Fock space to include up to one $\theta$-boson in the intermediate state, which reduces the   Schr\"odinger equation  to 
\begin{eqnarray} \label{eigeneq}
\begin{bmatrix} H_{22} & H_{23}   \\  H_{32} & H_{33} \end{bmatrix}   \begin{bmatrix} |\phi^+\phi^-\rangle   \\  |\phi^+\phi^-\theta\rangle \end{bmatrix}
   &=& E \begin{bmatrix} |\phi^+\phi^-\rangle   \\  |\phi^+\phi^-\theta\rangle \end{bmatrix}. 
\end{eqnarray}
The three-particle sector can be formally eliminated, resulting  in an effective two-body equation, 
\begin{equation}\label{twobodyeigeneq}
 (E -H_{22})|\phi^+\phi^-\rangle   = H_{23} \frac{1}{E-H_{33}} H_{32}|  \phi^+\phi^-\rangle. 
\end{equation}

\subsection{Two-particle scattering in an infinite volume}

Before considering two-particle states on a torus, we will first review the scattering problem in   
 infinite-volume, with further details given in Appendix \ref{hamiltonian}. After eliminating the three-particle states $|\phi^+\phi^-\theta\rangle$ from the coupled system ({\it cf} Eq.(\ref{eigeneq}))   we are left with an equation for the center-of-mass frame momentum-space wavefunction, $\varphi_{JM}(\mathbf{q})$, which is a product of a radial wavefunction depending on the magnitude of the relative 3-momentum, $q=|{\bf q}|$,  and the spherical harmonic  of definite angular momentum, $(J,M)$, 
\begin{equation}\label{lippman}
	\varphi_{JM}(\mathbf{q}) = \frac{1}{\sqrt{s} - 2\sqrt{\mathbf{q}^2 + m^2}} \int\! \!\frac{d^3\mathbf{k}}{(2\pi)^3}\, V(\mathbf{q},\mathbf{k}) \,\varphi_{JM}(\mathbf{k}).
\end{equation}
Here, $E=\sqrt{s}$ is the energy of the two-particle system in the center-of-mass frame. The non-local potential, 
$V(\mathbf{q},\mathbf{k}) $, induced by $\theta$-exchange is given explicitly in Eq. (\ref{potential}). 
Expressing this equation in coordinate space via a Fourier transform gives
\begin{equation}
\psi_{JM}(\mathbf{r}) = \int\!\! d^3\mathbf{r}'\, G_0(\mathbf{r}-\mathbf{r}'; \sqrt{s}) \int\!\! d^3\mathbf{z}\, \tilde{V}(\mathbf{r}', -\mathbf{z})\, \psi_{JM}(\mathbf{z}), \label{lipp}
\end{equation}
where the free Green's function is given by
\begin{equation}
G_0(\mathbf{r}-\mathbf{r}'; \sqrt{s})  = \int\!\!\frac{d^3\mathbf{q}}{(2\pi)^3}\, \frac{e^{i\mathbf{q}\cdot (\mathbf{r}  - \mathbf{r}')   }}{\sqrt{s} - 2\sqrt{\mathbf{q}^2 - m^2}}. \label{free_green}
\end{equation}
The wavefunction satisfies a relativistic Schr\"odinger equation,
\begin{equation} 
\left( \sqrt{s} - 2\sqrt{-\nabla^2 + m^2} \right) \psi_{JM}(\mathbf{r}) = \int\!\! d^3\mathbf{z}\, \tilde{V}(\mathbf{r}, -\mathbf{z})\, \psi_{JM}(\mathbf{z}).\label{schro}
\end{equation}
While Eq.(\ref{lipp}) was derived in the context of a particular model, our subsequent derivation only requires the general form of the relativistic Lippmann-Schwinger equation.
The asymptotic component of the two-body wavefunction relevant to scattering is given by the large 
distance behavior of the  Green's function. Evaluating the integral in Eq.(\ref{free_green}) 
  ({\it cf.} Appendix \ref{hamiltonian}), we find 
\begin{align}\label{green_free}
G_0&(\mathbf{r}; \sqrt{s} = 2\sqrt{k^2+m^2}) = \nonumber \\ &-\frac{\sqrt{s}}{2}\, \frac{e^{ikr}}{4\pi r}
- \frac{1}{r}\int_m^\infty \!\!\frac{\rho\,d\rho}{(2\pi)^2} \,\sqrt{\rho^2 - m^2} \frac{e^{-\rho r}}{k^2 + \rho^2},  
\end{align}
with the first term on the right hand side dominating as $r \to \infty$. 
For a potential $\tilde{V}$ which falls at large separations, the solution to Eq (\ref{schro}) outside the range of the 
 interaction is given by 
\begin{align}
\psi_{JM}(\mathbf{r}) 
	&\to \frac{\sqrt{s}}{2m} i^J \, Y_{JM}(\hat{\mathbf{r}}) \left[ 4\pi  \, j_J(kr) + 
		i k\, f_J(k) \, h^+_J(kr) \right], \label{free_scat}	
		\end{align}
where $f_J(k)$ is the partial wave scattering amplitude, 
\begin{equation}
f_J(k) = -\frac{m}{i^{J}} \!\int\!\! d^3\mathbf{r}' d^3\mathbf{z} \, j_J(kr') \, Y^*_{JM}(\hat{\mathbf{r}}')\, \tilde{V}(\mathbf{r}',-\mathbf{z})\, \psi_{JM}(\mathbf{z}), \label{scat_amp}
\end{equation}
which up to the inelastic threshold can be parameterized in terms of a single real momentum-dependent parameter, the scattering phase-shift, $\delta_J(k)$, as
\begin{equation}
f_J(k) = \frac{4\pi}{k} e^{i\delta_J} \sin \delta_J. \nonumber
\end{equation}

\subsection{Two-particle scattering on a torus}\label{torus}
  
Now we consider the theory in a cubic box of volume $V=L^3$, with periodic boundary conditions. In Eq. (\ref{lipp}) we split the integral over $\mathbf{r}'$ into a sum of integrals over a set of boxes labelled  by the integers $\mathbf{n}$ representing the location of one of its corners,
\begin{align}
\psi^{(L)}_{JM}(\mathbf{r}) &= \sum_{\mathbf{n} \in \mathbb{Z}^3}     \int_{L^3}\!\! d^3\mathbf{r}'\, G_0(\mathbf{r}-\mathbf{r}' - \mathbf{n}L; \sqrt{s}) \nonumber \\
&\times \int\!\! d^3\mathbf{z}'\, \tilde{V}(\mathbf{r}' +  \mathbf{n}L, -\mathbf{z}'-  \mathbf{n}L )\, \psi^{(L)}_{JM}(\mathbf{z}'+  \mathbf{n}L). \label{cubic_lipp}
\end{align}
In general we can make the wavefunctions periodic up to a phase,
\begin{equation}
\psi^{(L)}_{JM}(\mathbf{z} + \mathbf{n} L) = e^{i \mathbf{Q}\cdot \mathbf{n} L} \psi^{(L)}_{JM}(\mathbf{z}), \nonumber
\end{equation}
where the Bloch wave-vector, $\mathbf{Q}$, is related to the total momentum of the two-particle system \cite{Gottlieb:1995} by $\mathbf{P} = 2 \gamma \,\mathbf{Q}$. $\gamma = \sqrt{s+\mathbf{P}^2}/\sqrt{s} $ is the Lorentz contraction factor that reduces the effective size of the box in the direction parallel to $\mathbf{P}$. Using the periodicity of the potential, $\tilde{V}(\mathbf{r}' + \mathbf{n} L , - \mathbf{z}'-\mathbf{n} L) = \tilde{V}(\mathbf{r}' , - \mathbf{z}')$, and the boundary condition on the wavefunction, we have
\begin{align}
\psi^{(L,\mathbf{Q})}_{JM}(\mathbf{r}) &=     \int_{L^3}\!\! d^3\mathbf{r}'\, G_{\mathbf{Q}} (\mathbf{r}-\mathbf{r}'; \sqrt{s}) \nonumber \\
&\quad\quad\times \int\!\! d^3\mathbf{z}\, \tilde{V}(\mathbf{r}' , -\mathbf{z})\, \psi^{(L,\mathbf{Q})}_{JM}(\mathbf{z}), \nonumber  
\end{align}
which is analogous to the infinite-volume equation, but with the  Green's function given by 
\begin{equation}
	G_\mathbf{Q} (\mathbf{r}-\mathbf{r}'; \sqrt{s}) = \sum_{\mathbf{n} \in \mathbb{Z}^3}  G_0(\mathbf{r}-\mathbf{r}' - \mathbf{n} L; \sqrt{s})\, e^{i \mathbf{Q}\cdot\mathbf{n} L}. \nonumber
\end{equation} 
Using the Poisson summation formula, $(2\pi)^{-3} \sum_{\mathbf{n} \in \mathbb{Z}^3}   e^{i \mathbf{Q}\cdot \mathbf{n} L} = L^{-3} \sum_{\mathbf{n} \in \mathbb{Z}^3} \delta\left(\mathbf{Q} + \tfrac{2\pi}{L} \mathbf{n} \right)$, 
we  obtain 
\begin{align}
	G_\mathbf{Q} (\mathbf{r}-\mathbf{r}'; \sqrt{s}) & =   \frac{1}{L^{3}} \sum_{\mathbf{q} \in P_\mathbf{Q}}  \frac{ e^{i \mathbf{q} \cdot (\mathbf{r} - \mathbf{r}')   }  }{ \sqrt{s} - 2 \sqrt{\mathbf{q}^2 + m^2}  } \nonumber \\
	 & \to  \frac{\sqrt{s}}{2} \frac{1}{L^3} \sum_{\mathbf{q} \in P_\mathbf{Q}}  \frac{e^{i\mathbf{q}\cdot(\mathbf{r}- \mathbf{r}')} }{k^2 - \mathbf{q}^2}, \nonumber
\end{align}       
where $P_\mathbf{Q} = \{ \mathbf{q} \in \mathbb{R}^3 | \, \mathbf{q} = \tfrac{2\pi}{L}\mathbf{n} + \mathbf{Q},\; \text{for} \; \mathbf{n} \in \mathbb{Z}^3 \}$, and where we have retained only the leading term in the limit $L  \gg |{\bf r} - {\bf r'}|  \gg m^{-1}$.  
Finally, expanding Eq.(\ref{cubic_lipp}) for $r \gg r'$ and using the definition of the scattering amplitude, 
Eq. (\ref{scat_amp}) we can express the wavefunction as
\begin{align}
\psi^{(L,\mathbf{Q})}_{JM}(\mathbf{r}) & \to    \frac{\sqrt{s}}{2m} (-k)\, i^J f_J(k)  
	\sum_{J'M'} Y_{J'M'}(\hat{\mathbf{r}}) \nonumber \\
	& \times     \left[ \delta_{JM, J'M'}\, n_{J'}(kr) - \mathcal{M}^{(\mathbf{Q})}_{JM,J'M'}(k) \; j_{J'}(kr) \right].  \label{vol_scat}
\end{align} 
The residual sum over all angular momenta reflects the broken rotational invariance induced by the finite cubic volume, with the volume-dependent matrix elements ${\cal M}$  given in Appendix \ref{appendgreen}. 
In the infinite-volume case, the most general solution of the relativistic Schr\"odinger equation, Eq.(\ref{schro}), outside the range of the potential is $\sum_{JM} c_{JM} \,\psi_{JM}(\mathbf{r})$ for $\psi_{JM}(\mathbf{r})$ given by Eq. (\ref{free_scat}). Correspondingly in finite-volume, the most general solution is given by $\sum_{JM} c_{JM} \,\psi^{(L,\mathbf{Q})}_{JM}(\mathbf{r})$ for $\psi^{(L,\mathbf{Q})}_{JM}(\mathbf{r})$ given by Eq.(\ref{vol_scat}).  
Matching the two wavefunctions at a fixed $r$, larger than the range of the interaction, we obtain 
\begin{align} 
 &\sum_{JM} c_{JM} Y_{JM}(\hat{\bf r}) \, i^J \left[ 4\pi\, j_J(kr) + ik \,f_J(kr) \,h^+_J(kr) \right] \nonumber \\
&\quad= -\!\!\!\sum_{JM,J'M'} \!\!c_{JM} \,  i^J k \, f_J(k) \, Y_{J'M'}(\hat {\bf r}) \nonumber \\
 &\quad\quad\times \left[ \delta_{JM,J'M'}\,  n_{J'}(kr) - {\cal M}^{(\bf Q)}_{JM,J'M'} (k) \, j_{J'}(kr) \right], \nonumber
\end{align} 
which has a non-trivial, $c_{JM} \ne 0$, solution provided 
\begin{equation}
\det \left[  \delta_{JM,J'M'} \,\cot \delta_J(k) - \mathcal{M}^{(\mathbf{Q})}_{JM,J'M'}(k)    \right] = 0. \label{luescher}
\end{equation}
This condition expresses the relationship between the asymptotic behavior of the two-particle wavefunction on a torus, expressed through the matrix elements $ \mathcal{M}^{(\mathbf{ Q})}_{JM, J'M' }$, and the effect of the interaction on the wavefunction determined by the phase shifts,   $ \delta_{J}$. In practice for a given set of elastic phase-shifts, $\delta_J(k)$, it determines a discrete spectrum of states in a finite volume.

The analysis presented here can be  generalized to an arbitrary shaped box. In general the three edges of box are spanned by three arbitrary vectors $\mathbf{ L}_{1,2,3}$.  The volume of the cube $L^{3}$ is replaced by $(\mathbf{ L}_{1} \times \mathbf{ L}_{2} )\cdot \mathbf{ L}_{3}$ and  the vector $\mathbf{ n} L $  by $\sum_{i=1,2,3} n_{i} \mathbf{ L}_{i}$, $n_{i} \in \mathbb{Z}$. Finally the momentum \mbox{$\mathbf{ q}= 2\pi \mathbf{ n}/L, \mathbf{ n} \in \mathbb{Z}^{3}$} 
 is replaced by generalized momentum   \mbox{$ 2\pi   \sum_{i=1,2,3}  n_{i} (\mathbf{ L}_{j} \times \mathbf{ L}_{k} )/|(\mathbf{ L}_{1} \times \mathbf{ L}_{2} )\cdot \mathbf{ L}_{3}| ,\,  n_{i} \in \mathbb{Z}$ }, where indices $(i,j,k)$ follow  the cyclic  permutation.  Such a generalization has to be considered when 
  using the moving center-of-mass frame since the symmetric shape of a cubic box in the rest frame is deformed 
   due to Lorentz contraction \cite{Gottlieb:1995}. In this case if \mbox{$\mathbf{ P} = 2\pi \mathbf{ d}/L,\mathbf{ d} \in \mathbb{Z}^{3}$}, is the center-of-mass momentum, the volume of the box becomes  $\gamma L^{3}$, and the vectors $\mathbf{ n} L $  and  $2\pi \mathbf{ n}/L$  are replaced by  $ \gamma  \mathbf{ n}  L$ and  $ 2\pi\gamma^{-1}   \mathbf{ n}/ L  $,  respectively (using the notation defined in \cite{Gottlieb:1995}). With these substitutions  and  the relation $\mathbf{ P} =2 \gamma \mathbf{ Q}$, our definition of the matrix elements $ \mathcal{M}^{(\mathbf{Q})}_{JM,J'M'}$   becomes identical 
     to the matrix elements $ M^{\mathbf{d}}_{lm,l'm'}$ in Eq.(89) of \cite{Gottlieb:1995}.

       Typically, as discussed in \cite{Lusher:1991},  for the low-energy region that we are interested in, higher partial waves become progressively smaller and can be ignored, so that  the partial wave basis  can be truncated at a certain maximal 
angular momentum  $J_{\mathrm{max}}$. For a finite-volume with cubic boundaries, the continuous rotation symmetry is reduced to the little-group of allowed cubic rotations that leave the centre-of-mass momentum invariant - the matrices in Eq.(\ref{luescher}) become block-diagonal if subduced  according to  the irreducible representations of  these little groups. Details of 
subduction in general moving frames can be found in \cite{Christopher_subduction:2012}.

  
\section{Coupled channel scattering in    finite volume }  \label{coupledchannel}

  We extend the model of the previous section to include additional two-particle asymptotic states, 
 by adding another species of charged bosons, $\sigma^{\pm}$, which also couples to the force carrier, $\theta$, into the  Lagrangian.   We can obtain coupled equations for the two-particle states, $|\phi^+\phi^-\rangle$ and $|\sigma^+\sigma^-\rangle$, by  eliminating states featuring three or more particles and obtain a two-channel Schr\"odinger equation, 
   \begin{align}
\big|\phi^+\phi^-\big\rangle &= \frac{1}{E-H_{\phi}^{(0)} } \left[ V_{\phi\phi} \big|\phi^+\phi^-\big\rangle + V_{\phi\sigma} \big|\sigma^+\sigma^-\big\rangle\right], \nonumber \\
\big|\sigma^+\sigma^- \big\rangle &= \frac{1}{E-H_{\sigma}^{(0)} } \left[ V_{\sigma\phi} \big|\phi^+\phi^-\big\rangle  + V_{\sigma\sigma} \big|\sigma^+\sigma^- \big\rangle \right],  \label{ch}
\end{align}  
where $H_{\phi}^{(0)}$, $H_{\sigma}^{(0)}$ are the one-particle operators and $V_{\phi\phi}, V_{\phi\sigma}, V_{\sigma\phi}, V_{\sigma\sigma}$ are effective interactions (two-body operators) generated by the reduction
 to the two-particle subspace. From Eq.(\ref{ch}), for the channel  wavefunctions, $ \psi^{\alpha = \phi,\sigma}_{JM}(\mathbf{r})  \equiv \langle \mathbf{ r} | \alpha, JM \rangle$, we obtain 
\begin{align}
	 \psi^\alpha_{JM}(\mathbf{r}) &= \int\!\!d^3\mathbf{r}'\, G^\alpha_0(\mathbf{r}-\mathbf{r}'; \sqrt{s}) \nonumber \\
	&\quad \times \sum_{\beta}   \int\!\! d^3 \mathbf{z}\, \tilde{V}_{\alpha\beta}(\mathbf{r}', -\mathbf{z})\, \psi^\beta_{JM}(\mathbf{z}). \nonumber
\end{align}
The coupled-channel scattering amplitudes can be defined by
\begin{align}
	f_J^{\alpha\beta}(s) = -\frac{m_\alpha}{i^J} \int\!\! d^3 \mathbf{r}' d^3 \mathbf{z}\; &j_J(k_\alpha r') \, Y_{JM}^*(\hat{\mathbf{r}}')\nonumber \\
	\, &\times\tilde{V}_{\alpha\beta}(\mathbf{r}', -\mathbf{z})\, \psi^\beta_{JM}(\mathbf{z}), \label{f_amp}
\end{align}
where $k_\alpha = \sqrt{s/4- m_\alpha^2}$ is the  magnitude of the relative momentum in the center-of-mass frame of the two particles in  channel $\alpha$. By analogy to the single channel case, the asymptotic wavefunction in channel $\alpha$ is given by 
\begin{align}
	 \psi^\alpha_{JM}(\mathbf{r})& \to \frac{\sqrt{s}}{2m_\alpha} Y_{JM}(\hat{\mathbf{r}})\, i^J \nonumber \\
	&\times\left[  \,4\pi \, j_J(k_\alpha r) + i k_\alpha \, h^+_J(k_\alpha r)   \,{\textstyle \sum_\beta} f^{\alpha\beta}_J(s) \right]. \label{inf_coupled}
\end{align}
Extending the one-channel analysis of the asymptotic states in finite-volume to the two-channel system, 
 one obtains, 
\begin{align}
	& \,\psi^{\alpha (L,\mathbf{Q})}_{JM}(\mathbf{r}) \to \frac{\sqrt{s}}{2m_\alpha}(\text{-}k_\alpha) \, i^J 
	 \sum_{J'M'} Y_{J'M'}(\hat{\mathbf{r}}) \left[\textstyle{\sum_\beta f^{\alpha\beta}_J(s)}\right] \nonumber \\
	&\quad\quad\times\left[ \delta_{JM,J'M'}\, n_{J'}(k_\alpha r) - \mathcal{M}^{(\mathbf{Q})}_{JM,J'M'}(k_\alpha)\, j_{J'}(k_\alpha r)  \right]. \nonumber \\ \label{box_coupled}
\end{align}
Matching the wavefunctions in finite-volume, Eq. (\ref{box_coupled}), to the wavefunctions in infinite volume, Eq. (\ref{inf_coupled}),  we get a determinant condition
\begin{widetext}
\begin{align}
	\det\left[ \delta_{JM,J'M'}\,  \delta_{\alpha,\beta} \frac{4\pi}{k_{\alpha}}  \frac{1}{ f^{\alpha\alpha}_J}+  \left [ \,  i \, \delta_{JM,J'M'}\,   - \mathcal{M}^{(\mathbf{Q})}_{JM,J'M'}(k_\alpha)\,  \right ] \, \frac{f^{\alpha\beta}_J }{ f^{\alpha\alpha}_J} \,   \right]=0.  \label{coupscattrep}
\end{align}
The derivation logically extends to any number of scattering channels. Expressing the scattering amplitudes using $t$-matrix elements, $t_{\alpha \beta}^{(J)}(s) \equiv \tfrac{\sqrt{s}}{8\pi} f^{\alpha\beta}_J(s)$, and introducing the phase-space for channel $\alpha$ by $\rho_\alpha(s) = \tfrac{2 k_\alpha}{\sqrt{s}}$, we can write the condition as
\begin{equation}
0 = \det \left[  \delta_{JM,J'M'} \delta_{\alpha ,\beta} + i \rho_\alpha(s) \,  t^{(J)}_{\alpha \beta}(s)  \Big( \delta_{JM,J'M'} + i \mathcal{M}^{(\bf{Q})}_{JM,J'M'}(k_\alpha) \Big) \right], \label{det-t}
\end{equation}
or, alternatively in a form which expresses the effect of unitarity more directly as,
\begin{equation}
0 = \det \left[  \delta_{JM,J'M'} \Big( \big[ t^{(J)}(s) \big]^{-1}_{\alpha \beta} + i \rho_\alpha(s) \delta_{\alpha,\beta} \Big)  - \delta_{\alpha,\beta} \, \rho_\alpha(s)\,  \mathcal{M}^{(\bf{Q})}_{JM,J'M'}(k_\alpha) \Big) \right]. \label{det_unit}
\end{equation}
The multichannel unitarity condition can be expressed as $\mathrm{Im} \big[ t^{(J)}(s) \big]^{-1}_{\alpha \beta} = -\delta_{\alpha \beta} \rho_\alpha(s) \Theta( s - s_\mathrm{thr.}^{(\alpha)} )$, and thus, since $\rho_\alpha(s)$ becomes pure imaginary below threshold, the first term in Eqn \ref{det_unit} is always real. 

The form presented in Eqn \ref{det-t} can be shown to be equivalent to that presented in \cite{Briceno:2012yi} and \cite{Hansen:2012tf}. Their expressions include an additional phase of $i^{J-J'}$ in front of $\mathcal{M}$, but the effect of this phase is always cancelled in the determinant.

Reflecting the remaining symmetry of a cube in flight, there is in fact one determinant condition for each irreducible representation of the symmetry group. As presented in \cite{Jo_scatt:2012}, \cite{Christopher_subduction:2012} these conditions can be obtained by subduction, the result being conditions
\begin{equation}
0 = \det \left[  \delta_{JJ'} \delta_{nn'}  \Big( \big[ t^{(J)}(s) \big]^{-1}_{\alpha \beta} + i \rho_\alpha(s) \delta_{\alpha,\beta} \Big)  - \delta_{\alpha,\beta} \, \rho_\alpha(s)\,  \mathcal{M}^{(\mathbf{Q}, \Lambda)}_{Jn, J'n'}(k_\alpha) \Big) \right], \label{det_subduced}
\end{equation}
where the $\Lambda$-irrep subduced $\mathcal{M}^{(\mathbf{Q}, \Lambda)}_{Jn, J'n'}(k_\alpha)$ is as defined in Eqn (28) of \cite{Jo_scatt:2012}. The angular-momentum space is defined by the various embeddings, $n$, of spin-$J$ into the irrep $\Lambda$. If the subduction conventions in \cite{Jo_scatt:2012} are followed, for unitary $t$-matrices, the conditions are purely real at all real energies.

The two-channel  scattering matrix $f^{\alpha\beta}$ can be conventionally parameterizated in terms of
 two scattering phase-shifts, $\delta^\alpha_J(s)$ ($\alpha=\phi,\sigma$),  and an inelasticity, $\eta_J(s)$,
  representing the  fraction of flux exchanged between the two channels, 
\begin{align}
	f^{\alpha\alpha}_J(s) = \frac{4\pi}{k_\alpha} \cdot\frac{\eta_J \,e^{2i\delta^\alpha_J} - 1}{2i}; \quad\quad f^{\alpha\beta}_J(s) = \frac{4\pi}{\sqrt{k_\alpha k_\beta}} \cdot\frac{\sqrt{1-\eta_J^2} \, e^{i(\delta^\alpha_J + \delta^\beta_J)} }{2},\label{coupscattparam} 
\end{align}
so that the determinant condition can be written
\begin{align}
\det \begin{bmatrix}  \delta_{JM,J'M'}\, \cot \Delta^\phi_J - \mathcal{M}^{(\mathbf{Q})}_{JM,J'M'}(k_\phi)  & \sqrt{\tfrac{k_\phi}{k_\sigma}} \left[ i \,\delta_{JM,J'M'} - \mathcal{M}^{(\mathbf{Q})}_{JM,J'M'}(k_\phi) \right] \frac{\sqrt{1-\eta_J^2}\, e^{i \Delta_J^\sigma}}{2 \eta_J \sin \Delta_J^\phi}\\
	\sqrt{\tfrac{k_\sigma}{k_\phi}} \left[ i\, \delta_{JM,J'M'} - \mathcal{M}^{(\mathbf{Q})}_{JM,J'M'}(k_\sigma) \right] \frac{\sqrt{1-\eta_J^2}\, e^{i \Delta_J^\phi}}{2 \eta_J \sin \Delta_J^\sigma}     &     \delta_{JM,J'M'}\, \cot \Delta^\sigma_J  - \mathcal{M}^{(\mathbf{Q})}_{JM,J'M'}(k_\sigma) 
    \end{bmatrix} = 0, \label{cc_luescher}	
\end{align}
\end{widetext}\color{black} 
where $\Delta_J^\alpha(s) \equiv \delta^\alpha_J(s) - \tfrac{i}{2} \log \eta_J(s)$. One can show that this result is equivalent to  Eq.(4.14) in  \cite{Liu:2005}.

The determinant conditions presented above for coupled-channels scattering provide only one equation, at each finite-volume energy, for many unknowns. At low energies we may be justified in only considering the effect of the lowest contributing partial-wave, but even then there are multiple unknowns. For example in the case of two-channel scattering in $S$-wave we require three parameters to describe the $t$-matrix at each energy which might be the two phase-shifts and inelasticity, $\delta^\phi_0,\, \delta^\sigma_0,\, \eta_0$. Hence additional constraints need to be imposed to obtain a unique solution. As an example, in \cite{Doring:2011} unitarized chiral perturbation was used to constrain amplitude parameters at low energies. In the next section we explore some strategies for extraction of the scattering amplitude from finite-volume spectra in the context of an analytical parametrization of the amplitude.

It is worth noting here that in a finite volume, kinematically closed channels can play a role in determining the spectrum. Examining Eqn \ref{det-t}, we see that the behavior of $\mathcal{M}^{(\mathbf{Q})}_{JM,J'M'}(k)$, analytically continued below threshold (to imaginary $k$) must be considered. These matrix elements decay rapidly below threshold, such that the effect of a kinematically closed channel on the finite-volume spectrum is only felt in a limited energy region below the kinematic threshold. For example if $k = i \kappa$, 
\begin{equation}\mathcal{M}^{(\mathbf{Q})}_{00,00}(i \kappa) = i - \frac{i}{\kappa} \sum_{\substack{ \mathbf{ n} \in \mathbb{Z}^{3} \\ {\mathbf{ n}  \neq 0}} }\frac{e^{-\kappa |\gamma \mathbf{ n} L|}}{  |\gamma \mathbf{ n} L|} e^{i \mathbf{ Q} \cdot \gamma \mathbf{ n} L},  \nonumber
\end{equation}
so that far below threshold, or in very large volume, $\mathcal{M} \to i$, and thus in Eqn \ref{det-t}, the effect of this channel is removed. However this closed channel will have an effect in a region just below the threshold, for $\gamma \kappa \lesssim L^{-1}$. Hence we are required in practice to analytically continue scattering amplitudes below thresholds in order to determine the finite-volume spectrum. An example of this is presented in \cite{Torres:2012} for the case of $S$-wave $\pi \Sigma,\overline{K} N$ scattering.


\section{A toy model of resonant coupled-channel scattering in finite-volume}\label{toycoup}

In order to explore possible strategies for extracting coupled-channel scattering amplitudes from the discrete  finite-volume spectra emerging from lattice QCD computations, we  consider a simple model of two-channel $S$-wave scattering. The model is based on resonance-dominated scattering and satisfies multi-channel unitarity and the analytical properties required of partial wave amplitudes. With an explicit model for scattering amplitudes we can solve Eq. (\ref{det-t})  to obtain finite-volume spectra of states as a function of the volume ($V = L^3$) 
 and total momentum of the center-of-mass $\mathbf{P} = 2\pi \mathbf{d}/L,\; \mathbf{d} \in \mathbb{Z}^3$ ($\mathbf{P} = 2\gamma{\bf Q}$).   We then use this spectrum as pseudo-data representing a hypothetical lattice QCD simulation and attempt to reproduce the input model. 

\subsection{Analytic model of two-channel scattering}\label{toyinf}
We consider a model in which a single $S$-wave  resonance coupled to both scattering channels interferes with a non-resonant background. The two-channel scattering amplitude is parametrized in terms of a  $K$-matrix, 
\begin{equation}
	K_{\alpha\beta}(s) = \frac{g_\alpha g_\beta}{M^2 - s} + \gamma^{(0)}_{\alpha\beta} + \gamma^{(1)}_{\alpha\beta} \, s + \ldots,\label{Kmatrix}
\end{equation}
which is related to the $t$-matrix by 
\begin{equation}
	\left[t^{-1}(s)\right]_{\alpha\beta} = \left[K^{-1}(s) \right]_{\alpha\beta} + \delta_{\alpha\beta}\, I_\alpha(s), \nonumber
\end{equation}
and to the scattering amplitude defined in  Eq. (\ref{f_amp})  by $t^{\alpha\beta}_J(s) = \sqrt{s}  f^{\alpha\beta}_J(s)/8\pi$ with ($\alpha=\phi,\sigma$). Here  $I_\alpha(s)$ is the Chew-Mandelstam form  \cite{Basdevant:1977}, 
\begin{equation}
	I_\alpha(s) = I_\alpha(0) - \frac{s}{\pi} \int_{4m_\alpha^2}^\infty \!\!\!\! ds' \sqrt{1-\frac{4m_\alpha^2}{s'} } \frac{1}{(s'-s)s'}, \nonumber
\end{equation}
whose imaginary part above threshold, {\it i.e.} for $s> 4m_\alpha^2$, is the negative of the phase-space, $\mathrm{Im}\big[ I_\alpha(s) \big] = - \rho_\alpha(s)$. This form ensures the unitarity of the amplitude and provides a smooth transition across the kinematic threshold. We have opted to subtract the integral once, and it is convenient to choose $I_\alpha(0)$ such that $\text{Re}\, I_\alpha(M^2) = 0$ so that we have an amplitude which for real $s$ near $M^2$ is close to the Breit-Wigner form with mass $M$. 
The   $t$-matrix thus constructed is an analytical function in the complex $s$-plane with the discontinuity across the right-hand cut determined by unitarity.

\begin{figure}
\includegraphics[width=0.5\textwidth 
]{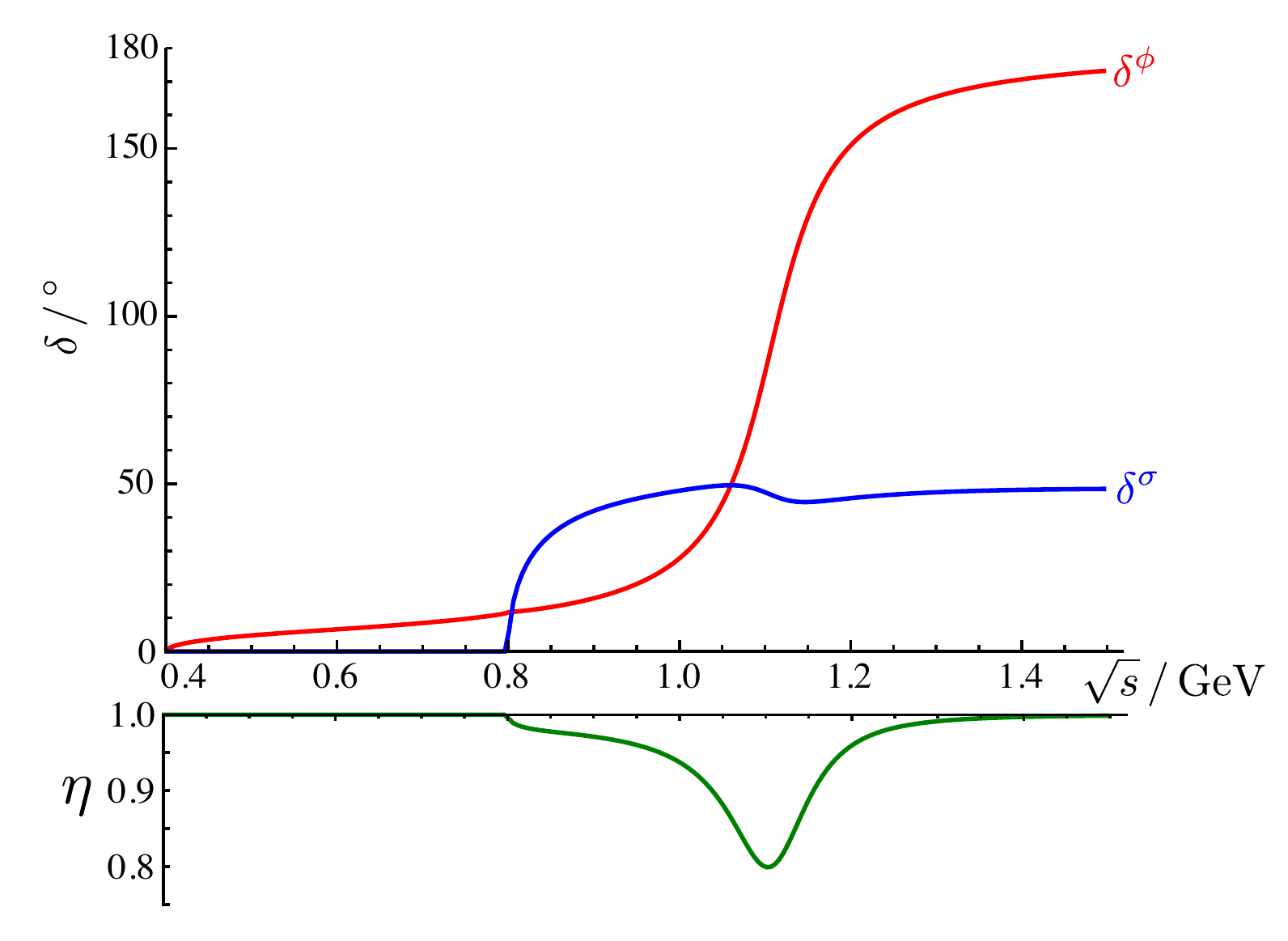}
\caption{Phase-shifts and inelasticity for the model defined in the text.
\label{fig:Kmatrix}}
\end{figure}

With the following choice of parameters, 
\begin{align}
	m_\phi = 0.2 \,\mathrm{GeV},\;\; m_\sigma = 0.4 \,\mathrm{GeV} \nonumber \\
	M = 1.1 \,\mathrm{GeV},\; g_\phi = 0.35\,\mathrm{GeV},\; g_\sigma = 0.2\,\mathrm{GeV} \nonumber \\
	\gamma^{(n)}_{\phi\phi} = \gamma^{(n)}_{\phi\sigma} = 0, \; \gamma^{(0)}_{\sigma\sigma} = 0.7,\; \gamma^{(1)}_{\sigma\sigma} = 0.7 \,\mathrm{GeV}^{-2}, \; \gamma^{(n>1)}_{\sigma\sigma} = 0, \nonumber
\end{align}
we obtain the phase-shifts and inelasticity shown in Fig.~\ref{fig:Kmatrix}. The parameters have been chosen in such a way that there is a narrow resonance 
near $\sqrt{s}= 1.1\,\mathrm{GeV}$. 
 It is usual to analyse scattering in terms of the most relevant singularities of the $t$-matrix on the nearby unphysical sheets. Poles on unphysical sheets are often identified with hadron resonances. In this case the four sheets ($\mathsf{sheet\, I}$ is the physical sheet) can be defined by
\begin{equation}
\left[ t_\mathsf{sheet}^{-1}(s)\right]_{\alpha\beta} = \begin{cases}
	\left[t_\mathsf{I}^{-1}(s)\right]_{\alpha\beta}  & \mathsf{sheet\,  I} \\ 
	\left[t_\mathsf{I}^{-1}(s)\right]_{\alpha\beta} + 2i\sqrt{1- \tfrac{4m_\phi^2}{s}} \delta_{\alpha \phi} & \mathsf{sheet \, II} \\ 
	\left[t_\mathsf{I}^{-1}(s)\right]_{\alpha\beta} + 2i\sqrt{1- \tfrac{4m_\alpha^2}{s}} \delta_{\alpha \beta} & \mathsf{sheet \, III} \\ 
	\left[t_\mathsf{I}^{-1}(s)\right]_{\alpha\beta} + 2i\sqrt{1- \tfrac{4m_\sigma^2}{s}} \delta_{\alpha \sigma} & \mathsf{sheet \, IV} 
\end{cases} . \nonumber
\end{equation}
The model amplitude has a single pole on the lower half-plane\footnote{and a conjugate pole on the upper half-plane} of each of sheets $\mathsf{II}$ and $\mathsf{III}$,  with the $t$-matrix in the neighbourhood of the pole at $s_0$ behaving like,
\begin{equation}
	\left[t_\mathsf{sheet}(s \to s_0)\right]_{\alpha\beta} \to \frac{c_\alpha c_\beta}{s_0-s}, \nonumber
\end{equation}
with 
\begin{align*}
 &\; \sqrt{s_0} = \big( 1.1067 - \tfrac{i}{2}0.0961 \big)\, \mathrm{GeV} \\
 & \; c_\phi = \big( 0.3585\, \mathrm{GeV} \big)\, e^{-i 0.0023\pi} ,\; c_\sigma = \big( 0.1367  \, \mathrm{GeV} \big) \, e^{-i0.237\pi}
 \end{align*} 
on $\mathsf{sheet\,II}$ and 
\begin{align*}
 &\; \sqrt{s_0} = \big( 1.1088 - \tfrac{i}{2}0.1195 \big)  \, \mathrm{GeV} \\
  & \; c_\phi = \big(0.3573 \, \mathrm{GeV}\big)\, e^{+i 0.0026\pi} ,\; c_\sigma = \big( 0.1391 \, \mathrm{GeV}\big)\, e^{+i0.297\pi}
\end{align*}
on $\mathsf{sheet\,III}$. 
Our aim is to use the finite-volume spectrum determined on a set of volumes and total momenta, $\mathbf{P}$, 
 to reproduce the pole positions of this scattering amplitude.


\subsection{Finite-volume spectrum}\label{spectrum}
\begin{figure*}
\includegraphics[width=1.06\textwidth 
]{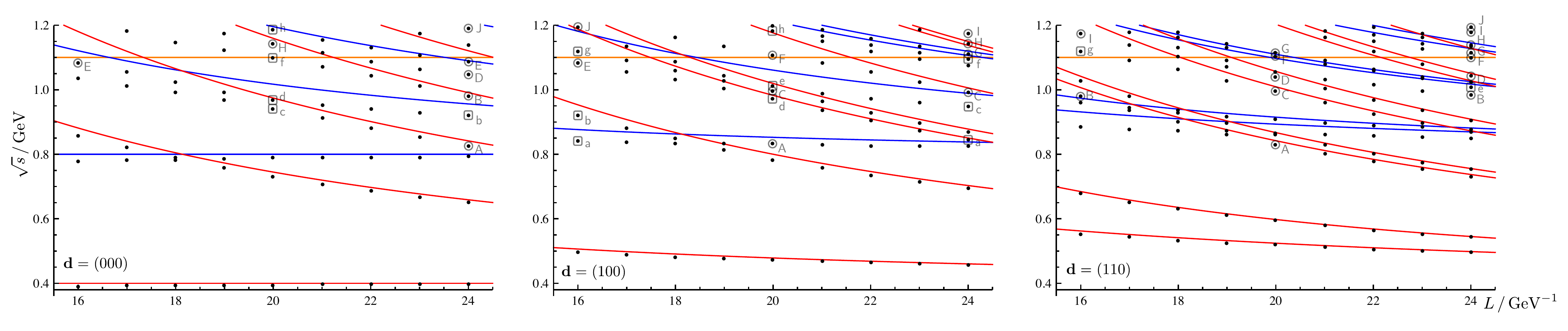}
\caption{Finite-volume spectra for the $K$-matrix model described in Sec.~\ref{toyinf}. Black dots indicate the spectrum obtained  by solving  Eq. (\ref{det-t}).  Red and blue curves represent the energy of a non-interacting pair of mesons $(\alpha=\phi,\sigma$) $ \left[ \left(\sqrt{m_{\alpha}^{2} + \mathbf{ k}^{2}_{1}}  +\sqrt{m_{\alpha}^{2} + \mathbf{ k}^{2}_{2}}  \right)^{2} - \mathbf{ P}^{2} \right]^{1/2} $, $\mathbf{ k}_{1} + \mathbf{ k}_{2}= \mathbf{ P}$ and $\mathbf{ k}= \frac{2 \pi}{L} \mathbf{ n}, \mathbf{ n} \in \mathbb{Z}^{3}$. The points labelled by letters are used as described in the text.
\label{fig:spectrum}}
\end{figure*}

The finite-volume spectrum corresponding to the model defined in the previous section can be obtained by solving Eq. (\ref{cc_luescher}) (or equivalently Eq. (\ref{det-t})), where for $E = \sqrt{s} < 2m_\sigma$ we require the analytic continuation of $t$-matrix elements featuring channel $\sigma$. Assuming that partial waves higher than $S$-wave are negligible reduces Eq. (\ref{cc_luescher}), for $E>2m_\sigma$, to, 
\begin{align}
	0 &= \Omega(\delta^\phi(E), \delta^\sigma(E), \eta(E); L, \mathbf{d}; E) \nonumber \\
	&= \eta \left[ \mathcal{M}_{\phi} - \mathcal{M}_\sigma \right] \sin \left( \delta^\phi - \delta^\sigma \right) \nonumber \\
	&\quad\quad+ \left[ \mathcal{M}_{\phi} +\mathcal{M}_\sigma \right] \sin \left( \delta^\phi + \delta^\sigma \right) \nonumber \\
	&\quad\quad-  \eta \left[1+ \mathcal{M}_{\phi} \, \mathcal{M}_\sigma \right] \cos \left( \delta^\phi - \delta^\sigma \right) \nonumber \\
	&\quad\quad-   \left[1- \mathcal{M}_{\phi} \, \mathcal{M}_\sigma \right] \cos \left( \delta^\phi + \delta^\sigma \right) , \label{condition}
\end{align}
where $\mathcal{M}_{\phi} \equiv \mathcal{M}^{(\mathbf{Q})}_{J=0\,M=0,J'=0\,M'=0}(k_\phi)$ with 
 a similar expression for $\mathcal{M}_{\sigma}$. $\mathbf{Q}$ is a function of 
 $\mathbf{d}$, $L$, $E$ as discussed in Sec.~\ref{torus}. 
 In Fig.\ref{fig:spectrum} we show the finite-volume spectrum\footnote{this would be the spectrum in irrep $A_1$ \cite{Jo_scatt:2012}} obtained by solving the determinant condition as a function of the volume in a region   $L=16-24 \mbox{ GeV}^{-1}$  (or $L=3.2- 4.7 \mbox{ fm}$).

\subsection{``Pointwise" estimation of scattering from finite-volume spectrum}\label{pointwise}

\begin{figure*}
\includegraphics[width=0.8\textwidth 
]{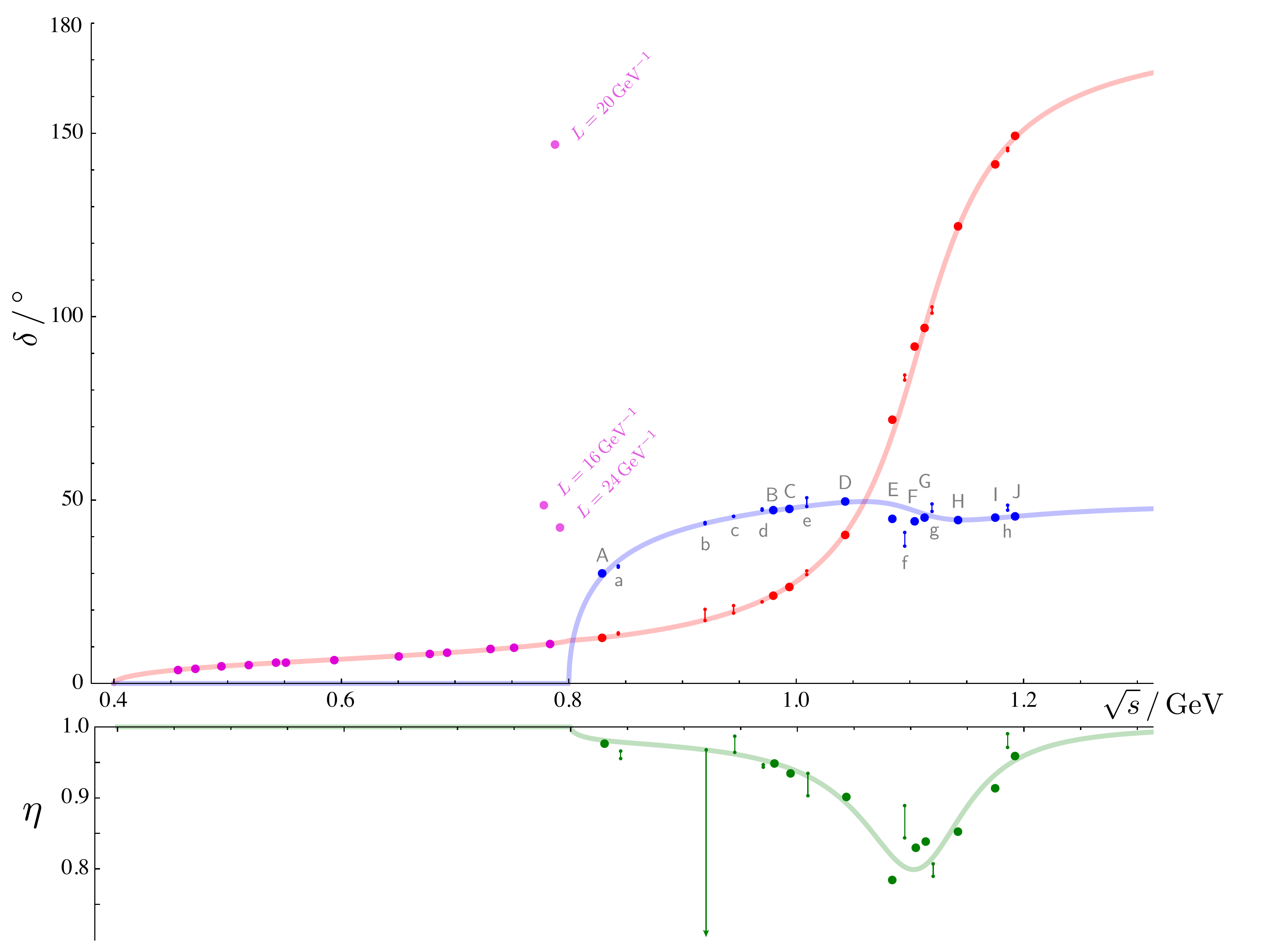}
\caption{``Pointwise" determination of the phase-shifts and inelasticity. $\delta^\phi$ points below the opening of the $\sigma\sigma$ threshold (in pink) are determined by solving the elastic relation, Eqn. (\ref{luescher}), ignoring the effect of the closed channel. The badly discrepant points just below the threshold are all within a momentum scale of $1/L$ of the threshold where the finite-volume effect of the closed-channel should not be neglected. 
Energies $\mathsf{A}-\mathsf{J}$ are determined from constrained three-level analysis, energies $\mathsf{a}-\mathsf{h}$ by interpolation in $\delta^\phi$ from two-level analysis. The light-colored curves show the exact input model.
\label{pointwise_results}}
\end{figure*}

In the region below $2m_\sigma$ the $\phi\phi$ scattering is elastic and it is tempting to use Eq (\ref{luescher}), completely ignoring the existence of the kinematically closed $\sigma\sigma$ channel. Doing so leads to the points shown in pink in Fig \ref{pointwise_results} which successfully reproduce the input model except in a region just below the threshold. As previously discussed, in a finite-volume the effect of a closed channel can be felt within a limited energy region immediately below the threshold. With this in mind, we must be careful not to use the simple elastic relation, Eq (\ref{luescher}), close to a threshold.

One approach to determining the phase-shifts and inelasticities  at discrete values of energy above $2m_\sigma$ is to locate multiple energy levels (in different volumes and/or different $\mathbf{d}$) that appear at approximately the same energy. As an example consider the three levels labeled $\mathsf{A}$ in Fig. \ref{fig:spectrum} which all lie within 3 MeV of $\sqrt{s_\mathsf{A}} = 830\,\mathrm{MeV}$. 
For the three levels we can build three independent copies of Eq. (\ref{cc_luescher}) which each feature approximately the same values of $\delta^\phi(s_\mathsf{A}), \delta^\sigma(s_\mathsf{A}), \eta(s_\mathsf{A})$, which can be determined by solving the set of simultaneous equations. Since the energies are not \textit{exactly} degenerate, there need not be an exact solution to the equations and hence we seek to find the solution which minimizes
\begin{align}
	\sum_{E(L,\mathbf{d})} \left| \Omega(\delta^\phi, \delta^\sigma, \eta; L, \mathbf{d}; E)\right|^2, \nonumber
\end{align}
with $\Omega$ defined in Eq. (\ref{condition}) and where the sum is over the three energy levels. For the levels $\mathsf{A}$, the obtained solution, as shown in Fig. \ref{pointwise_results}, is within $3\%$ of the exact  value of $\delta^\phi(s_\mathsf{A}), \delta^\sigma(s_\mathsf{A}), \eta(s_\mathsf{A})$. We emphasize that this procedure is not guaranteed to successfully reproduce the true scattering amplitudes - the $\mathcal{M}$ matrices can in some circumstances vary rather rapidly over a narrow energy region. 
  
 Within the energy region considered, $E=0.8-1.2\,\mathrm{GeV}$, considering only three volumes, $L=16,\, 20,\, 24\, \mathrm{GeV}^{-1}$, and three sets of center-of-mass momentum,  $\mathbf{d}=(000),\, (100),\, (110)$, we can isolate a number of such sets. These sets of three near-degenerate energy levels are 
  labelled $\mathsf{A}-\mathsf{J}$ in  Fig.~\ref{fig:spectrum}. In 
  Fig.~\ref{pointwise_results} the labels are shown on the plot of $\delta^\phi$ with the 
  corresponding solutions for $\delta^\sigma$ and $\eta$ marked by solid dots.

With these points alone, in Fig. \ref{pointwise_results} we see strong hints of a signal for resonant behavior in the $\delta^\phi$ phase shift. While obviously reasonably successful, this approach does not make optimal use of the finite-volume spectral information, by failing to use any energy level which does not have two near-degenerate partners. To use somewhat more of the discrete levels we might consider building a system of \textit{two} instances of Eq. (\ref{cc_luescher}) with one parameter from the set $\delta^\phi, \delta^\sigma, \eta$, estimated using interpolation between already determined values. 
   For example consider the two levels  near $1.009\mbox{ GeV}$  labeled $\mathsf{e}$ 
    in Fig. \ref{fig:spectrum}. Linear interpolation between the energies of the $\mathsf{C}$ and $\mathsf{D}$ points in 
    Fig.~\ref{pointwise_results} gives $\delta^\phi(1.009\,\mathrm{GeV}) = 30.7^\circ$. Using this value 
   and minimizing with respect to $\delta^\sigma, \eta$ at the energy corresponding to point $\mathsf{e}$ 
    we obtain $\delta^\sigma = 50.6^\circ, \eta = 0.903$. A spline interpolation using all the points  $\mathsf{A-J}$  yields $\delta^\phi(1.009\,\mathrm{GeV}) = 29.6^\circ$ which results in $\delta^\sigma = 48.1^\circ, \eta= 0.934$  at 
     the point  $\mathsf{e}$. 
In Fig. \ref{pointwise_results} we show the results for sets of two degenerate levels labeled $\mathsf{a}-\mathsf{h}$ from Fig. \ref{fig:spectrum}. In each case the range shown indicates limiting values obtained using   two 
  methods of interpolation.  
  Even though  in some cases
   there is a considerable sensitivity to the interpolation method, overall the points are in reasonable agreement with the  model input (solid light-colored curves).
 

\subsection{Parameterized estimation of scattering from finite-volume spectrum}\label{global_fit}

The previously discussed ``pointwise" strategy, while having the advantage of being largely model-independent, is reliant upon there being multiple energy levels which, through accident or design, are close to degenerate. Since it would be unusual to engineer lattice volumes purely for this purpose, and unusual in contemporary calculations to have such a high density of determined energy levels, it is appropriate to consider alternative methods of analysis. One such approach that makes full use of all determined levels, and which may require far fewer levels to be determined, involves parameterising the scattering amplitude and performing a minimisation to get the best description of the determined finite-volume spectrum by varying the parameters. In the current toy-model, even limited ``pointwise" analysis would suggest that the phase $\delta^\phi$ is rapidly rising and would indicate that a resonance could be present. In practical calculations (e.g.\cite{Jo:2010}, \cite{Dudek:2012xn}), the presence of a sharp meson resonance can also be indicated by large overlap onto fermion bilinear operators. By including a pole (as well as polynomial behaviour) in a $K$-matrix parameterisation we are likely to get rapid convergence to a solution with a pole in the $t$-matrix corresponding to the resonance. 

\begin{figure}
\includegraphics[width=0.35\textwidth]
{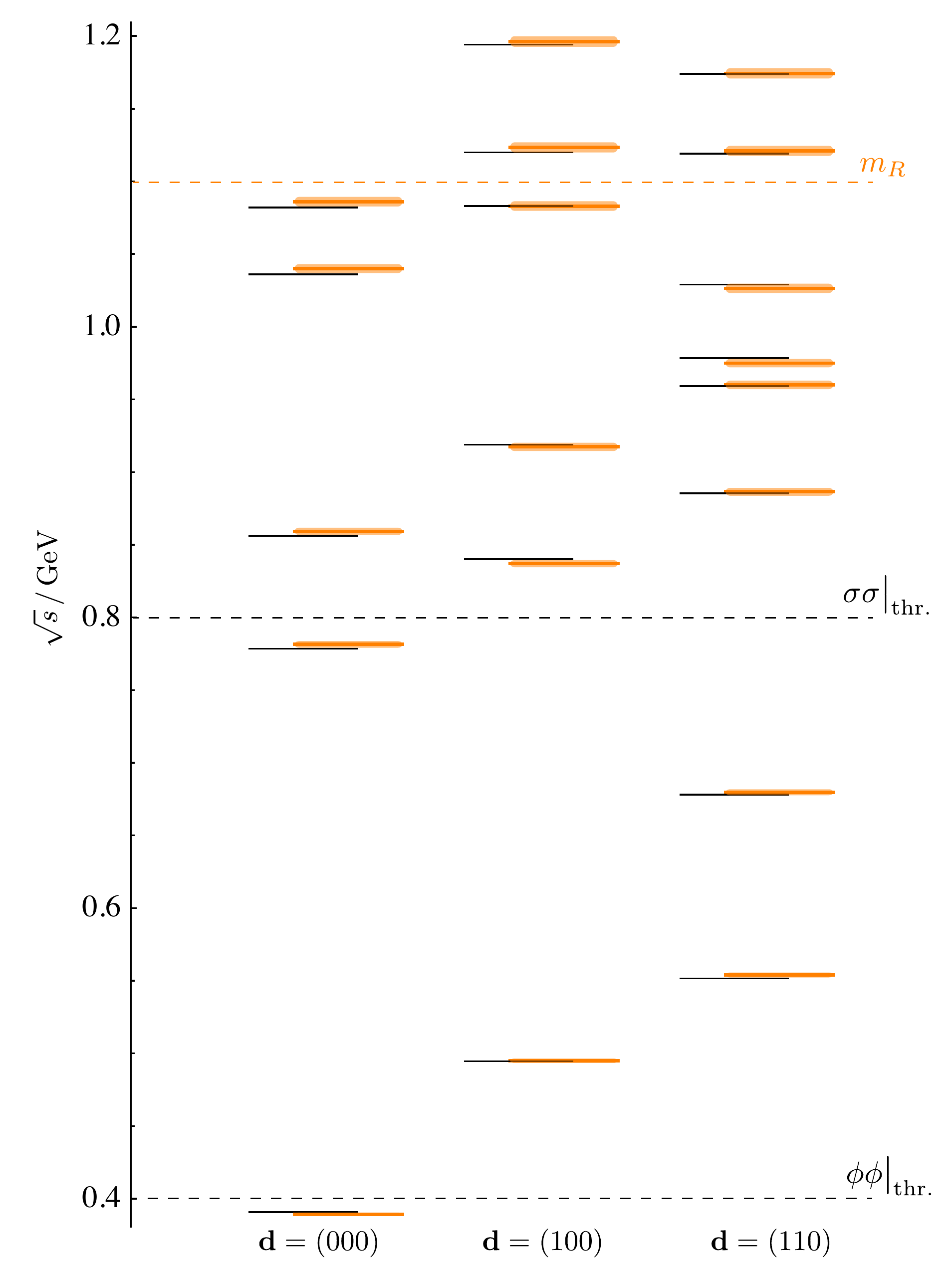}
\caption{$L = 16 \,\mathrm{GeV}^{-1}$. Orange rectangles: finite-volume spectra with 0.3\% noise. Black lines: exact finite-volume spectrum given in Fig.\ref{fig:spectrum}. Also shown the position of the thresholds and the $K$-matrix pole mass.
\label{fig:spectrum_noise}}
\end{figure}

We will take this opportunity to make the toy model a slightly more realistic simulation of an actual lattice QCD calculation by introducing statistical uncertainty on the energy level values. In recent work \cite{Jo_scatt:2011}, \cite{Jo_scatt:2012}, \cite{Dudek:2012xn}, the Hadron Spectrum Collaboration has obtained statistical errors on excited levels as small as 0.3\% and we will assume that this remains practical. For each energy level below $1.2$ GeV on a single volume $L=16\,\mathrm{GeV}^{-1} \sim 3.2\,\mathrm{fm}$ with $\mathbf{d}=(000),\,(100),\,(110)$, we randomly generated an ensemble by drawing from a distribution whose mean is the exact value given in Fig. \ref{fig:spectrum} and whose variance is chosen such that the ensemble has variance on the mean of 0.3\% of the mean value. The resulting spectrum is shown in Fig.  \ref{fig:spectrum_noise}.

\begin{figure*}
\includegraphics[width=0.99\textwidth 
]{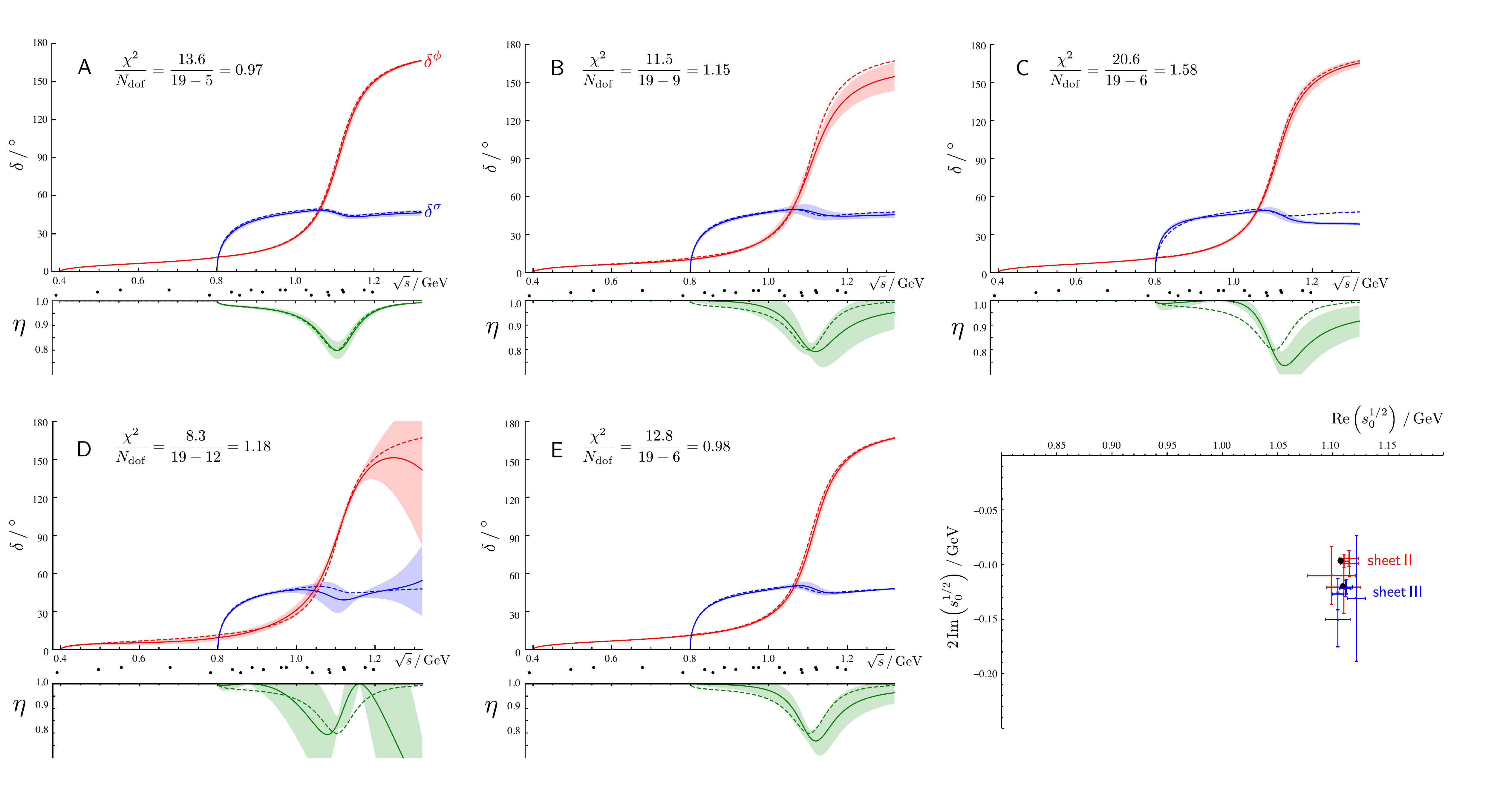}
\caption{
Best-fit phase-shifts and inelasticity for the parameterizations $\mathsf{A-E}$ as described in the text. Solid curves with error bands show the minimized solutions with the dashed curves showing the exact input. Black dots under the energy scale show the positions of the discrete energy levels used in the minimization. Bottom-right panel shows the determined position of a $t$-matrix pole on the unphysical sheets $\mathsf{II}$(red) and $\mathsf{III}$(blue) for the five parameterisations along with the exact input positions (black dots).
\label{fit_results}}
\end{figure*}

Parametrizing according to a form like that given in Eq. (\ref{Kmatrix}), we can minimise a function,
\begin{equation}
\chi^2\big(\{a_i\}\big) = \sum_{E_\mathfrak{n}(L,\mathbf{d})} 
\frac{ \left[ E_\mathfrak{n}(L,\mathbf{d}) - E^\text{det}_\mathfrak{n}(L,\mathbf{d}; \{a_i\}) \right]^2
}{\sigma(E_\mathfrak{n}(L,\mathbf{d}))^2}, \nonumber  
\end{equation}
(where $E^\text{det}$ are the solutions of Eq. (\ref{det-t})), by varying the parameters of the $K$-matrix parameterisation, $\{a_i\} = \{M, g_\phi, g_\sigma, \gamma^{(n)} \ldots \} $. In practical lattice QCD calculations, the $\chi^2$  can be trivially redefined to deal with correlated data by replacing the inverse diagonal variance ($1/\sigma^{2}$) by the inverse of the data covariance matrix.

In Fig.\ref{fit_results} we show the parameterized phase-shifts and inelasticity obtained using five model parameterisations: 
 
\begin{itemize}
\item $\mathsf{A}$: ``exact model",  which uses a 1st order polynomial in $s$ to describe the non-pole contribution to the $K$-matrix with $\gamma^{(0,1)}_{\phi \phi} =  \gamma^{(0,1)}_{\phi \sigma} \equiv 0$ [5 parameters]
\item $\mathsf{B}$: ``relaxed model", with a 1st order polynomial in all channels {\it i.e.} all  $\gamma^{(0,1)}$ free [9 parameters] 
\item $\mathsf{C}$: ``tight model", with 0th order of polynomials in all channels {\it i.e.}  $\gamma^{(1)} = 0$ [6 parameters]. 
\item $\mathsf{D}$: ``loose model", with 2nd order of polynomials in all channels {\it i.e.}  $\gamma^{(0,1,2)}$ free [12 parameters]. 
\item $\mathsf{E}$: ``two pole model", with no polynomial, but with a second $K$-matrix pole with independent variable couplings [6 parameters]. 
\end{itemize}
As one would expect, within statistical uncertainty,  parameterization $\mathsf{A}$ reproduces the input model.  
  Parameterisation $\mathsf{B}$, which is more flexible, also reproduces the input quite well over the energy region where data is given, albeit with a larger statistical uncertainty, but begins to show signs of deviation from the original $K$-matrix in the energy range outside of the fit region.  Parameterization $\mathsf{C}$ does not have sufficient flexibility to describe the complete energy dependence - while it does correctly reproduce the resonance shape in $\delta^\phi$ and the presence of a dip in $\eta$, the precise energy dependence of $\eta$ is not correct and it fails to describe $\delta^\sigma$ at high energies, away from the energy region where the pole dominates.  Parameterisation $\mathsf{D}$ introduces too much parameter-space freedom for the limited set of data points available. As such we see rapid energy variation that is not really required and a large degree of statistical uncertainty.  Parameterisation $\mathsf{E}$ shows that a precise knowledge of the form of the ``background" is not required to reproduce the energy dependence in a limited region - a second $K$-matrix pole (at higher energy) is able to mock-up the polynomial behavior away from the resonance pole quite well.
  
Our principal interest  lies in identifying resonances as poles in the complex-$s$ plane -- analytically continuing the fitted model amplitudes we find that all five have single poles on sheets $\mathsf{II}$ and $\mathsf{III}$ whose locations are in a rather good agreement with the input pole position (see Fig. \ref{fit_results}). The residues at the pole agree similarly. The statistical precision of the pole position determination typically decreases as we introduce more parameters into the model. The pleasing observation here is that in order to determine the position of a sharp resonance we do not need to have precise knowledge of the form of the energy-dependence of the ``background".

In summary, the ``pointwise'' strategy may provide a less model-dependent approach for extracting phase shifts and inelasticities, however, this method is limited by the number of points for which accidental degeneracies appear. Parameterizing the scattering amplitude allow us to make use of all measured energy levels, however we need to find suitable parameterizations. One strategy is to explore the ``pointwise" approach to find a crude guide to the energy dependence and then build parameterizations which are able to reproduce the obtained form. The parameterizations, which should respect certain constraints applicable to scattering amplitudes, can be made progressively more sophisticated in an effort to reduce the overall $\chi^2$ -- in this sense the approach is not dissimilar to what is done with real experimental data.

\section{Summary} \label{summary}

Using the Hamlitonian formalism applied to a model of interacting relativistic fields, we derived a generalized L\"uscher's formula \cite{Lusher:1991,Gottlieb:1995,Liu:2005,Doring:2011} for two-particle  scattering, in both the single- and coupled-channel systems, in moving frames. 

Our results are consistent with the ones obtained previously in \cite{Lusher:1991,Gottlieb:1995,Liu:2005,Doring:2011}. In the coupled-channel case we are challenged by the fact that, even for dominance of a single partial-wave, the system is underconstrained for determination of multiple-channel phase-shifts and inelasticities from a single determined finite-volume energy level. Using a toy model of two-channel $S$-wave scattering we demonstrated that it is possible to determine this information if multiple energy levels are determined.

 Two possible strategies for extracting information from discrete spectra of a coupled-channel system were discussed,  one approach utilizes the near degeneracy of energy levels in different volumes and total momenta of system, and another fits the discrete spectra  by parameterizating the scattering amplitudes with certain  numbers of parameters. These strategies may be useful  for the analysis of future lattice QCD data. In particular, the coupled-channel analysis has to be considered for the strongly coupled systems, for instance,   $\pi \pi, K \bar{K}$ and $ \eta \eta$ system in  $S$-wave.

\section{ACKNOWLEDGMENTS}
We thank  D. B. Renner and M. R. Pennington for useful discussions, and our colleagues within the Hadron Spectrum Collaboration for their continued assistance.  PG, RGE and JJD acknowledge support from U.S. Department of Energy contract DE-AC05-06OR23177, under which Jefferson Science Associates, LLC, manages and operates Jefferson Laboratory. JJD also acknowledges the support of the U.S. Department of Energy Early Career award contract DE-SC0006765. APS acknowledges the support of the U.S. Department of Energy  grant under contract DE-FG0287ER40365.

\appendix


\section{Relativistic Lippmann-Schwinger equation from Hamiltonian formalism}\label{hamiltonian}

Following the method presented in \cite{Glazek:1993,Bakker:2003}, we   treat the relativistic dynamics of  particle scattering  in  the Hamiltonian formalism approach.  We  start from the covariant Lagrangian  in Eq.(\ref{Lagrangian}) and choose to quantize the field operators in the instant form \cite{Dirac:1949}; the construction of generators of the Poincar\'e group can be done in a standard  way in quantum field theory. In principle,  one needs to solve eigenstate equations $\hat{H}|\Psi\rangle=E|\Psi\rangle$ on a instant quantization plane, where $|\Psi\rangle$ denotes the Poincar\'e covariant state vector spanning the complete Fock space. We  truncate the Fock space up to three-body states,  assuming that this is sufficient to describe low-energy physics. Thus, the eigenstate equations reduce to a matrix equation given in Eq.(\ref{eigeneq}).  Eliminating the three-body sector, we end up with the relativistic Schr\"odinger-like equation for a two-body system given in Eq.(\ref{twobodyeigeneq}). For simplicity, we have assumed that the two charged scalars scattering have equal mass, however, the conclusion of this work can be generalized to non-equal masses case as well (c.f. \cite{Fu:2012,Prelovsek:2012}).

We choose the center-of-mass frame of the many-body system to construct multiple-particle states $|JM\rangle$  having total spin $J$ and spin projection $M$.   
The two-particle state  $|\phi^{+} \phi^{-} \rangle$ in the center-of-mass frame is given by
 \begin{align}
\big|\phi^{+} \phi^{-};J M\big\rangle  &= 2 \sqrt{s} \int\!\! \frac{d^{3} \mathbf{ p}_{1}}{(2\pi)^{3}  2 E_{p_{1}} }  \frac{d^{3} \mathbf{ p}_{2}}{(2\pi)^{3}  2 E_{p_{2}} }   \delta^{3} (\mathbf{ p}_{1} + \mathbf{ p}_{2}   ) \nonumber \\
&\quad\quad\quad\quad\times   (2\pi)^{3}    \varphi^{(2)}_{JM}(\mathbf{ p}_{1},\mathbf{ p}_{2}) \, a^{\dag}_{     \mathbf{ p}_{1}  } b^{\dag}_{   \mathbf{ p}_{2} } \,  \big|0\big\rangle  ,\nonumber
\end{align}
where $\mathbf{ p}_{i}$ is the momentum of the $i$-th particle and $\sqrt{s}$ is the invariant mass of the two-particle system.   $  \varphi^{(2)}_{JM}(\mathbf{ p}_{1},\mathbf{ p}_{2}) $ is the wavefunction of the two-particle system, describing the momentum distribution of the two particles. 

Similarly,  the three-particle state   $|\phi^{+} \phi^{-} \theta \rangle$ is given by
\begin{align}
 \big|\phi^{+} \phi^{-} \theta;J M  \big\rangle = 2 \sqrt{s} &\int \!\!\! \frac{d^{3} \mathbf{ p}_{1}}{(2\pi)^{3}  2 E_{p_{1}} }  \frac{d^{3} \mathbf{ p}_{2}}{(2\pi)^{3}  2 E_{p_{2}} }  \frac{d^{3} \mathbf{ p}_{3}}{(2\pi)^{3}  2 E_{p_{3}} }  \nonumber \\
&\quad\times    (2\pi)^{3} \delta^{3} (\mathbf{ p}_{1} + \mathbf{ p}_{2} + \mathbf{ p}_{3} )  \nonumber \\
&\quad\times \varphi^{(3)}_{JM}(\mathbf{ p}_{1},\mathbf{ p}_{2},\mathbf{ p}_{3})  \, a^{\dag}_{  \mathbf{ p}_{1} } b^{\dag}_{\mathbf{ p}_{2} } d^{\dag}_{\mathbf{ p}_{3}} \big|0\big\rangle  ,  \nonumber 
\end{align}
where  $  \varphi^{(3)}_{JM}(\mathbf{ p}_{1},\mathbf{ p}_{2},\mathbf{ p}_{3}) $ is the wavefunction of the three-particle system. The wavefunctions are normalized so that the normalization of states is  $\langle JM|JM\rangle =  2 \sqrt{s}\,  (2\pi)^{3} \, \delta^{3}(\mathbf{ 0})$.

It is straightforward to evaluate the matrix elements of the eigenstate equations Eq.(\ref{eigeneq})  and to get coupled equations for the wavefunctions 
\begin{widetext}
\begin{eqnarray} 
 &&  \varphi^{(2)}_{JM}(\mathbf{ q})    =   \frac{g}{\sqrt{s}-2 \sqrt{  \mathbf{ q}^{2}    +m^{2}}    }   \int\!\!\!    \frac{d^{3} \mathbf{ k}'}{(2\pi)^{3}  2  \sqrt{  \mathbf{ k'}^{2}    +\mu^{2}}   }          \left[   \frac{  \varphi^{(3)}_{JM}(\mathbf{ q}+ \tfrac{1}{2}\mathbf{ k'},\mathbf{ k'})}{   2  \sqrt{  (  \mathbf{ q}  + \mathbf{ k'})^{2}  +m^{2}} }  +   \frac{      \varphi^{(3)}_{JM}(\mathbf{ q}   - \tfrac{1}{2}\mathbf{ k'},\mathbf{ k'})}{   2  \sqrt{  (  \mathbf{ q} - \mathbf{ k'})^{2}  +m^{2}} } \right ]  ,      \nonumber  \\
 &&   \varphi^{(3)}_{JM}(\mathbf{ q}, \mathbf{ k})   = g\;   \frac{    \frac{1}{2\sqrt{( \frac{1}{2}\mathbf{ k}- \mathbf{ q} )^{2} +m^{2}}}  \,  \varphi^{(2)}_{JM}(\mathbf{ q} - \frac{1}{2} \mathbf{ k}) + \frac{1}{2\sqrt{( \frac{1}{2}\mathbf{ k}+ \mathbf{ q} )^{2} +m^{2}}}  \,  \varphi^{(2)}_{JM}(\mathbf{ q}+ \tfrac{1}{2}\mathbf{ k}) }{\sqrt{s}- \sqrt{ ( \frac{1}{2}\mathbf{ k} + \mathbf{ q} )^{2}   +m^{2}}  -   \sqrt{ ( \tfrac{1}{2}\mathbf{ k} - \mathbf{ q} )^{2}   +m^{2}  }    - \sqrt{\mathbf{ k}^{2}   +\mu^{2}  }}   ,    \nonumber
\end{eqnarray}
where  we have used a short-hand notation for wavefunctions $  \varphi^{(2)}_{JM}  (\mathbf{ q})$  and $  \varphi^{(3)}_{JM}  (\mathbf{ q}, \mathbf{ k})$, with arguments of relative momenta defined by $ \mathbf{ q}=\frac{1}{2}(\mathbf{ p}_{1}-\mathbf{ p}_{2}) , \mathbf{ k} =- \mathbf{ p}_{3}$. Eliminating the three-body wavefunction, we get a relativistic equation for the two-particle state $| \phi^{+} \phi^{-} \rangle$ with an effective non-local potential generated from the neutral scalar exchange between two charged scalars
\begin{eqnarray}\label{lpeq}
  \varphi_{JM}^{(2)}(\mathbf{ q})    =     \frac{1}{\sqrt{s}- 2\sqrt{  \mathbf{ q}^{2}    +m^{2}}    }   \int  \!\!  \frac{d^{3} \mathbf{ k}}{(2\pi)^{3}   }   \,        V(\mathbf{ q} ,\mathbf{ k} )  \,   \varphi_{JM}^{(2)}( \mathbf{ k}) , \nonumber 
  \end{eqnarray}
with
\begin{equation}\label{potential}
V(\mathbf{ q} ,\mathbf{ k} )   =  \frac{g^{2} }{4}\frac{1}{1-\Sigma(\mathbf{ q})}  \frac{1}{( \mathbf{ k}^{2}  +m^{2})} \frac{1}{   \sqrt{ (\mathbf{ k}-\mathbf{ q})^{2}    +\mu^{2}} }     \frac{1   }{\sqrt{s}- \sqrt{  \mathbf{ k}^{2}   +m^{2}}  -   \sqrt{    \mathbf{ q}^{2}   +m^{2}  }    - \sqrt{  ( \mathbf{ k}-\mathbf{ q})^{2}    +\mu^{2}  }} ,   
\end{equation}
where the self-energy contribution is, 
\begin{align} 
 \Sigma(\mathbf{ q})  =   \frac{1}{\sqrt{s}- 2\sqrt{  \mathbf{ q}^{2}    +m^{2}}    } \frac{g^{2} }{4} \; \int  \!\!\!  \frac{d^{3} \mathbf{ k}'}{(2\pi)^{3}       }     &   
   \frac{1   }{\sqrt{s}- \sqrt{  \mathbf{ k'}^{2}   +m^{2}}  -   \sqrt{  \mathbf{ q}^{2}   +m^{2}  }    - \sqrt{(\mathbf{ k'} - \mathbf{ q} )^{2}   +\mu^{2}  }}   \nonumber \\
  &\times \frac{1}{\sqrt{ \mathbf{ q}^{2} +m^{2}}}  \frac{  1 }{     \sqrt{    \mathbf{ k'}^{2}  +m^{2}} }  \frac{1}{\sqrt{  (\mathbf{ k'}-\mathbf{ q})^{2}    +\mu^{2}}} .      \nonumber
    \end{align}
 \end{widetext}

\begin{figure}
\begin{center}
\includegraphics[width=0.35 \textwidth ]{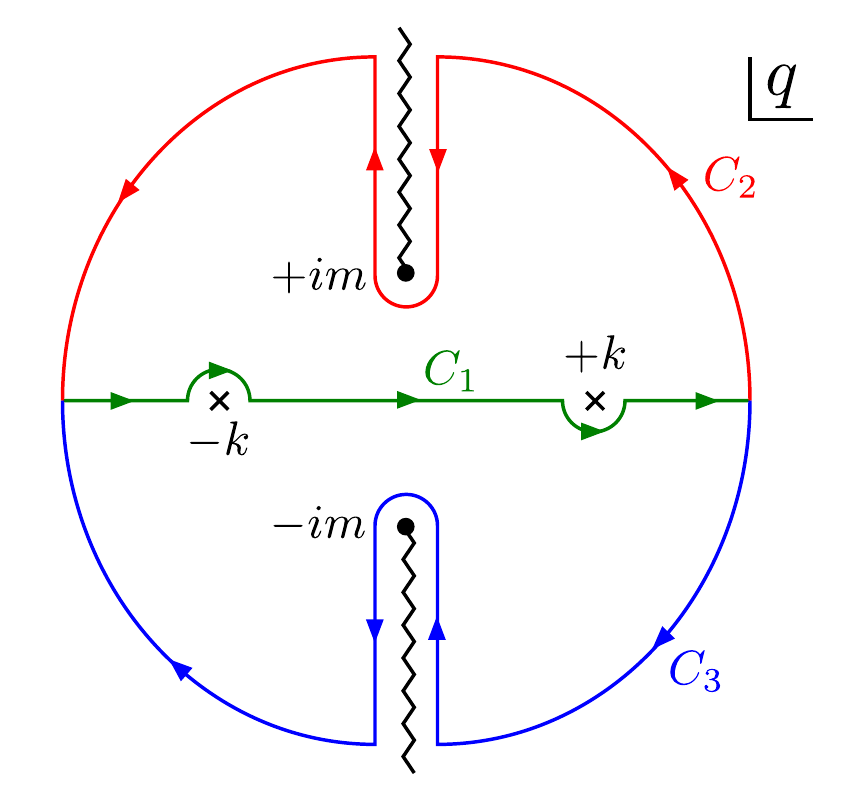}  
\caption{Integration contours and singularities of the free Green's function in Eq.(\ref{contour_green_ang}) on the complex $q$ plane.
\label{green_cut}}
\end{center}
\end{figure}

 In coordinate space, the wave-equation becomes
 \begin{eqnarray}\label{lpeq_coordinate}
\psi_{JM}(\mathbf{ r})=   \int \!\! d^{3} \mathbf{ r}'     \,  G_{0}(\mathbf{ r} -\mathbf{ r'},\sqrt{s})       \int \!\! d^{3} \mathbf{ z}  \,  \widetilde{ V}(\mathbf{ r'} , -\mathbf{ z} )   \,   \psi_{JM}( \mathbf{ z})    , \nonumber        
   \end{eqnarray}
   where $\psi_{JM}(\mathbf{ r})$ and $  \widetilde{ V}(\mathbf{ r'} , -\mathbf{ z} )$ are the Fourier transforms of the momentum-space two-particle wavefunction and effective potential, respectively. The free Green's function is given in Eq.(\ref{green_free}). Performing the angular integral, the free Green's function reads    
 \begin{eqnarray}\label{contour_green_ang}
G_{0}(\mathbf{ r}  ,\sqrt{s}) =  \frac{1}{ 2 i r}  \int \limits_{-\infty}^{\infty} \!\! \frac{ q d q }{(2\pi)^{2}} \frac{ e^{i qr} -e^{-i qr} }{  \sqrt{s}  - 2\sqrt{ q^{2}   + m^{2}}       }  ,    
   \end{eqnarray}   
which has the following singularities in the complex $q$ plane: two poles on the real axis, $q=\pm k$, and two branch cuts on the imaginary axis   $\pm \left [im, i \infty \right]$, see Fig.\ref{green_cut}. We choose the contour $C_{1}+C_{2}$ to include the pole at $q=k$ and circle around the cut $ \left [im, i \infty \right]$ on the upper half-plane for first term with factor  $e^{i kr}$ and choose the contour $C_{1}+C_{3}$ to include the pole at $q=-k$ and circle around the cut  $\left[-im, -i \infty \right]$ on the lower half-plane for the second term with factor  $e^{-i kr}$, as shown in  Fig.\ref{green_cut}.  The contour integral leads to
\begin{align} 
G_{0}(\mathbf{ r}  ,\sqrt{s}) & = -   \frac{\sqrt{s}}{2}   \frac{e^{i k  r } }{ 4\pi  r}  
- \frac{1}{r}  \int \limits_{m}^{ \infty }  \!\!  \frac{\rho d \rho }{(2\pi)^{2}}      \sqrt{   \rho^{2}-m^{2}}       \frac{ e^{ -\rho r}  }{   k^{2}+\rho^{2}     }     ,  \nonumber
   \end{align}   
 where $k = \tfrac{1}{2}\sqrt{s- 4 m^{2}}$ is the momentum of either particle in the rest frame of the two-particle system. The first term on the right hand side comes from the poles at $q=\pm k$ and is proportional to the usual non-relativistic Green's function which oscillates over the path of propagation.  The second term comes from the contribution of the discontinuity crossing the branch cuts at $\pm \left [im, i \infty \right]$ - it decays exponentially over the propagation. 
          Expanding $\sqrt{\rho^2 - m^2} = \rho \left(1- \mathcal{O}\big(\frac{m^{2}}{\rho^{2}} \big)   \right)$,  
at large separations,  the free Green's function may be approximated by
\begin{align}\label{green_free_expand}
G_0(\mathbf{r}; \sqrt{s} ) \approx -\frac{\sqrt{s}}{2}\, \frac{1}{4\pi r} \left[ e^{ikr}  + \frac{2}{\pi} \frac{e^{-mr}}{r \sqrt{s} } \right],
\end{align}
 and therefore the exponential decaying term can be dropped in the limit $r\gg m^{-1}$.
      
\section{Expansion of Green's function and regularization of expansion coefficients}\label{appendgreen}   
 We start from the expansion of the Green's function  
\begin{align}
\frac{1}{L^{3}}& \sum_{ \mathbf{ q} \in\mathbf{ P}_{\mathbf{ Q}}  }   \frac{e^{i  \mathbf{ q} \cdot  \mathbf{ r}  } }{ k^{2}- \mathbf{ q}^{2}       } \nonumber \\
  &= \frac{k}{4\pi} \, n_{0}(kr) - \sum_{j m_{j}} g^{(\mathbf{ Q})}_{j m_{j}}\!(k) \, j_{j}(kr) \, Y_{j m_{j}}(\mathbf{ \hat{r}}),  \nonumber
\end{align}
where the summation of $\mathbf{ q}$ runs over  $\mathbf{ P}_{\mathbf{ Q}} = \{ \mathbf{ q} \in \mathbb{R}^{3}| \mathbf{ q}= \frac{2\pi}{L}\mathbf{ n} + \mathbf{ Q}, \mbox{ for }   \mathbf{ n}\in \mathbb{Z}^{3} \} $.
 The expansion coefficients are given by \cite{Lusher:1991}
\begin{eqnarray}\label{exp_coef}
 g^{(\mathbf{ Q})}_{j m_{j}}\!(k) = \frac{4\pi}{L^{3}} \sum_{   \mathbf{ q} \in \mathbf{ P}_{\mathbf{ Q}}}   \! i^{j} \,  \frac{q^{j}}{k^{j}} \,  \frac{Y_{j m_{j}}^{*} \!(\mathbf{ \hat{q}}) }{   \mathbf{ q}^{2}  -k^{2} }   -    \frac{\delta_{j 0} \, \delta_{m_{j} 0}}{\sqrt{4\pi} }  \left. \frac{1}{r}\right|_{r \rightarrow 0}.   \nonumber  \\
\end{eqnarray}
 Note that the definition of  $n_{j}(x) $ in this work differs from the definition in \cite{Lusher:1991} by a overall negative sign.

Using the identities 
\begin{align}
    j_{j}& \!\big(  k| \mathbf{ r}-\mathbf{ r'} |  \big) \,    Y_{jm_{j}}( \widehat{\mathbf{ r}-\mathbf{ r'}} ) \nonumber \\
 &\stackrel{r'<r}{=} \sqrt{ 4\pi} \sum_{\substack{l m_{l} \\ l' m_{l'}}}  i^{l -l'-j} \, j_{l} (kr) \, j_{l'} (kr') \, Y_{l m_{l}} (\mathbf{ \hat{r}})   \, Y^{*}_{l' m_{l'}} (\mathbf{ \hat{r}'}) \nonumber \\
&\hspace{2cm}\times \sqrt{\tfrac{(2j+1)(2l'+1)}{ 2l+1}} \, \langle j m_{j}; l' m_{l'} | l m_{l}\rangle    \, \langle j 0 ; l' 0| l0\rangle, \nonumber
\end{align}
and
   \begin{align}
\frac{k}{4\pi} \, n_{0} \! \big(k| \mathbf{ r}-\mathbf{ r'} | \big) \stackrel{r'<r}{=}   k  \sum_{l m_{l}  }   n_{l} (kr) \,  j_{l} (kr') \, Y_{l m_{l}} (\mathbf{ \hat{r}})    \, Y^{*}_{l m_{l}} (\mathbf{ \hat{r}}') , \nonumber  
\end{align}
we also  have
\begin{align} 
 \frac{1}{L^{3}} &\!\sum_{  \mathbf{ q} \in \mathbf{ P}_{\mathbf{ Q}}}    \frac{e^{i  \mathbf{ q} \cdot  ( \mathbf{ r} -\mathbf{ r'}) } }{ k^{2}- \mathbf{ q}^{2}       }       \nonumber \\
&\stackrel{r'<r}{=}   \sum_{\substack{l m_{l} \\ l' m_{l'}}}  \left[ \delta_{l' m'_{l'}, l m_{l}}    \, n_{l} (kr)- \mathcal{M}^{(\mathbf{ Q})}_{l' m'_{l'}, l m_{l}} (k)   \, j_{l} (kr) \right] ,\nonumber \\
&\quad\quad\quad\quad\quad\times k\, j_{l'} (kr') \, Y_{l m_{l}} (\mathbf{ \hat{r}})    \, Y^{*}_{l' m_{l'}} (\mathbf{ \hat{r}'})  ,   \label{exp_green}
\end{align}
with
   \begin{align}\label{mat_element}
   \mathcal{M}&^{(\mathbf{ Q})}_{l' m'_{l'}, l m_{l}}(k)= \sum_{j m_{j}}   i^{l -l'-j}  \frac{\sqrt{4\pi}}{k}g^{(\mathbf{ Q})}_{j m_{j}} (k)  \nonumber \\
  &\times \sqrt{\tfrac{(2j+1)(2l'+1)}{ 2l+1}} \langle j m_{j}; l' m_{l'} | l m_{l}\rangle \langle j 0 ; l' 0| l0\rangle .
   \end{align}

If $\bf{Q}$ is identified with $\tfrac{1}{2\gamma} \bf{P}$ for two-equal-mass-particle scattering,   the generalization to two-unequal-mass-particle scattering  leads to $\frac{1}{2\gamma} \mathbf{ P} (1+ \frac{m_{1}^{2}- m_{2}^{2}}{E^{2}})$ \cite{Fu:2012,Prelovsek:2012},  these expressions are the same as those presented in \cite{Gottlieb:1995}, and the regularisation procedure outlined therein can be followed.   Once the Lorentz contraction  is considered (substitution of box volume $L^{3}$ and momentum $\frac{2\pi}{L} \mathbf{ n} , \mathbf{ n} \in \mathbb{Z}^{3}$   by $\gamma L^{3}$ and   $\frac{2\pi}{L}  \gamma^{-1}\mathbf{ n} $ respectively),  the function $ g^{(\mathbf{ Q})}_{j m_{j}}\!(k) $ is related to the Zeta function defined in Eq.(93) of \cite{Gottlieb:1995} by
\begin{eqnarray} 
 g^{(\mathbf{ Q}) }_{j m_{j}}\!(k) =     \frac{1}{\pi}     \frac{1}{\gamma L  }  \frac{i^{j}}{ \left(\frac{kL}{2\pi}\right)^{j}}   Z^{\mathbf{ d}*}_{j m_{j}} \left(1,\tfrac{kL}{2\pi} \right).      
\end{eqnarray}


\end{document}